# Reduction the secular solution to periodic solution in the generalized restricted three-body problem

**Elbaz I. Abouelmagd[1], M.E. Awad[2], E. M. A. Elzayat[3], Ibrahim A. Abbas[4]**

**[1,4]Mathematics Department, Faculty of Science and Arts (Khulais), King Abdulaziz University, Jeddah, Saudi Arabia**.

**[1]Email**: eabouelmagd@gmail.com or eabouelmagd@kau.edu.sa

**[2]Department of Space Science, Faculty of Science, Cairo University, Cairo, Egypt**
**[3]Mathematics Department, Faculty of Science and Arts (Khulais - Girls Branch), King Abdulaziz University, Jeddah, Saudi Arabia**.

## Abstract

The aim of the present work is to reduce the secular solution around the triangular equilibrium points to periodic solution in the frame work of the generalized restricted thee-body problem. This model is generalized in sense that both the primaries are oblate and radiating as well as the gravitational potential from a belt. We show that the linearized equation of motion of the infinitesimal body around the triangular equilibrium points has a secular solution when the value of mass ratio equals the critical mass value. Moreover, we reduce this solution to periodic solution, as well as some numerical and graphical investigations for the effects of the perturbed forces are introduced. This model can be used to examine the existence of a dust particle near the triangular points of an oblate and radiating binary stars system surrounded by a belt.

## 1 Introduction

The problem of three bodies in its most general form means that the three participating bodies are free to move in space and initially move in any given manner under the influence of a given force field. The significance of this problem in space dynamics will appear when the bodies move under the influence of their mutual gravitational attraction according to the Newtonian Law of gravitation. This law specifies that attractive forces between each pair of masses are inversely proportional to the squares of their distances and are proportional to the product masses of the respective particles.

A first consequence of this Law comes when two of the bodies approach each other such that the separation distance between them goes to zero and the force between them also comes to infinity. This circumstance is called double or triple collision according to whether two or three of the participating particles go to the same position in space at the same time. A second consequence of the force law, it follows that when one of the three participating particles is very smaller than the other two. In this situation the motion of the two larger particles will not be influenced by the smaller particle. This dynamical system is referred to as the restricted three-body problem. Therefore, if the motion of the smallest particle is found, we can determine the motion of the other two particles by setting the mass of the smaller particle as zero.

From the above discussion the restricted problem is an abstraction in the physical sense and an approximation in the mathematical sense. Since there is no effect for the smaller mass on the two larger particles unless its mass equals either zero or moves to infinity. Both of these cases reduce the restricted three-body problem to the problem of two bodies.

It is well known that there are five Lagrange solutions in the rotating coordinates system show up as five fixed points at which the infinitesimal body would be stationary if placed there with zero velocity. It is further known that in this rotating coordinates system the infinitesimal body may describe periodic orbits around Lagrange solutions which are referred to as Libration points.

Among the most fundamental questions about motion near libration points are those about the existence of periodic orbits and their stability. Periodic orbits obtain their significance to space mechanics when stable periodic orbits do exist.

They may be used as reference orbits. Furthermore, the determination of non-periodic orbits can be performed by perturbation analysis based on periodic orbits. In this section we will survey some results thus for obtained in the investigations of stability and periodicity.

The studies of many authors concerned to the existence of libration points, their stability and the periodic orbits in the framework of the restricted problem under the influence of the lack of sphericity, the photogravitational force, small perturbations in fictitious forces. Some of these works are introduced by Sharma (1987), Elipe and Lara (1997), Ishwar and Elipe (2001), Perdios (2001), Tsirogiannis et al (2006), Mittal et al., (2009), Singh and Begha (2011), Abouelmagd et al. (2013), Abouelmagd and El-Shaboury (2012) and Abouelmagd (2012, 2013a, 2013b).

Some researchers devoted their studied for exploring the families of asymmetric periodic orbits. Papadakis (2008) studied the asymmetric solutions of the restricted planar problem of three bodies. He explored numerically the families of asymmetric simple-periodic orbits. He also presented the evolution of these families covering the entire range of the mass parameter of the problem. Furthermore, he regularized the equations of motion of the problem using the Levi-Civita transformations to avoid the singularity due to binary collisions between the third body and one of the primaries. Symmetric relative periodic orbits in the isosceles three-body problem using theoretical and numerical approaches are studied by Shibayama and Yagasaki (2011). They proved that another family of symmetric relative periodic orbits is born from the circular Euler solution besides the elliptic Euler solutions. Their studies also showed that there exist infinitely many families of symmetric relative periodic orbits which are born from heteroclinic connections between triple collisions as well as planar periodic orbits with binary collisions.

Hou and Liu (2011) investigated that the collinear libration points of the real Earth–Moon system are not equilibrium points anymore due to various perturbations. They found special quasi-periodic orbits called dynamical substitutes under the assumption that the Moon's motion is quasi-periodic. In addition they computed the dynamical substitutes of the three collinear libration points in the real Earth–Moon system. In addition, Beevi and Sharma (2012) explored the effect of oblateness of Saturn on the periodic orbits and the regions of quasi-periodic motion around both the primaries in the Saturn-Titan system in the framework of planar circular restricted three-body problem. They studied the effect of oblateness on the location, nature and size of periodic and quasi-periodic orbits, using the numerical technique of Poincare surface of sections. They also showed that some of the periodic orbits change to quasi-periodic orbits due to the effect of oblateness and vice-versa.

Furthermore Abouelmagd and Sharaf (2013) studied and found these orbits around the libration points when the more massive primary is radiating and the smaller is an oblate spheroid. Their study included the effects of zonal harmonic parameters up to $10^{-6}$ of the main term.

The model of restricted three-body problem when the two primaries are oblate spheroids and radiating as well as the effect of gravitational potential from the belt are constructed by Singh and Taura (2013). They constructed the equations of motion, found the positions of the equilibrium points and examined their linear stability. They also established that, in addition to the usual five equilibrium points, there are two new collinear points $L_{n1}, L_{n2}$ due to the potential from the belt. They investigated that the collinear equilibrium points remain unstable, while the triangular points are stable for $\mu \in (0, \mu_c)$ and unstable for $\mu \in [\mu_c, 1/2]$, where $\mu_c$ is the critical mass influenced by the perturbed forces that are aforementioned.

In this work, we will follow Singh and Taura (2013) to find the secular solution around the triangular points in the restricted three-body problem and reduce this solution to periodic solution.

## 2 Model description

**2.1 Hypothesis:** We assume that $m_1$ and $m_2$ denote the masses of the more massive and the smaller primaries respectively and the mass of the infinitesimal body is $m$. Let us consider the same assumptions of Singh and Taura (2013). Both masses $m_1$ and $m_2$ have circular orbits around their common center of mass. Furthermore $m$ moves in orbital plane under their mutual gravitational fields. The sum of $m_1$ and $m_2$ is one where $\mu = m_2 / (m_1 + m_2)$ is the mass ratio and the distance between them also is taken as one. In addition the unit of time is chosen to make both the constant of gravitation and the unperturbed mean motion equals unity. Let the origin of the sidereal and the synodic coordinates be the common center of mass of the primaries and the synodic coordinates rotate with angular velocity $n$ in positive direction. Hence we can write $m_1 = 1 - \mu$ and $m_2 = \mu \le 1/2$, the coordinates of $m_1$, $m_2$ and $m$ in a synodic frame are $(-\mu, 0, 0)$, $(1 - \mu, 0, 0)$ and $(x, y, z)$ respectively, see Figure (1).

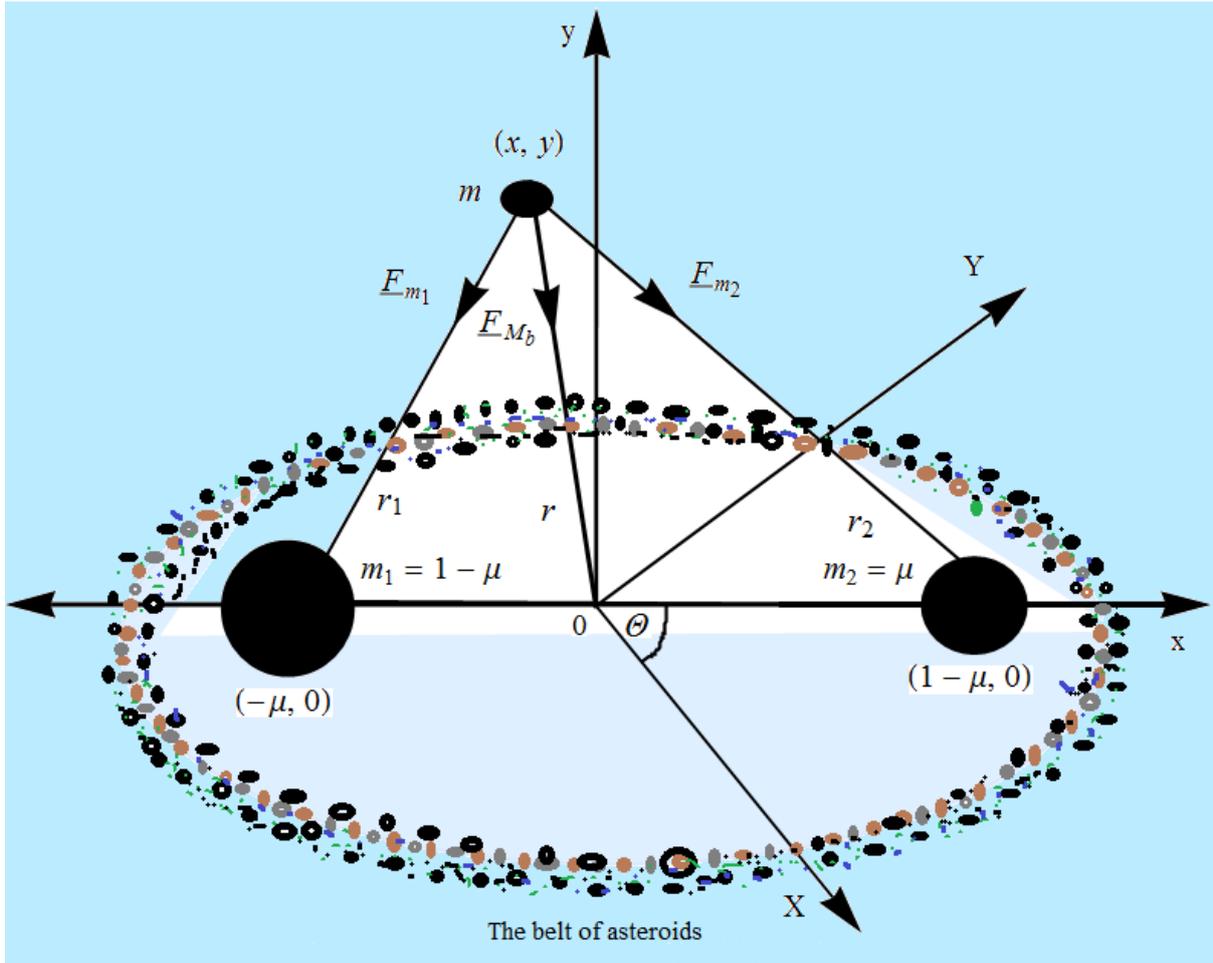

Fig. 1: The configuration of inertial XY and synodic x y coordinate frames of the restricted three-body problem when the primaries are surrounded by the belt of asteroids.

Now let the radiation parameter be $q_i = 1 - p_i$ and the oblateness coefficient is also $A_i$, $(i = 1, 2)$ for the bigger and smaller primaries respectively, where $0 < p_i \ll 1$, $p_i = F_{pi} / F_{gi}$ and $0 < A_i \ll 1$. Moreover, the potential due to the belt is $M_b / (r^2 + T^2)^{1/2}$ see Miyamoto and Nagai (1975) where $M_b$ is the total mass of the belt, $r$ is the radial distance of the infinitesimal body such that $r^2 = x^2 + y^2$, $T = a + b$, $a$ and $b$ are constants that characterize the density profile of the belt. Such that $a$ determine the flatness of the profile and is called the flatness parameter. While $b$ gives the size of the core of the density profile and is known as the core parameter. In the case $a = b = 0$, we obtain the potential of a point mass or spherical subject whose mass is $M_b$. Furthermore, the directions of forces ($\underline{F}_{m_1}$, $\underline{F}_{m_b}$ and $\underline{F}_{m_1}$) experienced by the mass $m$ can be shown as in Figure (1).

**2.2 The equation of motion:** We suppose that $m_1$ and $m_2$ move in $xy$ plane. The equations of motion of infinitesimal body given below as in Singh and Taura (2013)

$$\ddot{x} - 2n\dot{y} = \Omega_x \qquad (1.1)$$

$$\ddot{y} + 2n\dot{x} = \Omega_y \qquad (1.2)$$

here

$$\Omega = \left\{ \begin{array}{l} \dfrac{1}{2} n^2 [x^2 + y^2] + (1-\mu)q_1 [\dfrac{1}{r_1} + \dfrac{A_1}{2r_1^3}] \\[3mm] + \mu q_2 [\dfrac{1}{r_2} + \dfrac{A_2}{2r_2^3}] + \dfrac{M_b}{(r^2 + T^2)^{1/2}} \end{array} \right\} \qquad (2)$$

where $n$ the perturbed mean motion while $r_1$ and $r_2$ are distances of $m$ with respect to $m_1$ and $m_2$ respectively, that are given by

$$r_1^2 = \left( x + \mu \right)^2 + y^2 \qquad (3.1)$$

$$r_2^2 = \left( x + \mu - 1 \right)^2 + y^2 \qquad (3.2)$$

$$n^2 = 1 + \dfrac{3}{2}(A_1 + A_2) + \dfrac{2M_b r_c}{(r_c^2 + T^2)^{3/2}} \qquad (4)$$

If we multiply (1.1), (1.2) by $\dot{x}$ and $\dot{y}$ respectively and add them, we will get a perfect differential for $\Omega$ where $\Omega \equiv \Omega(x, y)$ is a function of $x, y$, hence after integrating we obtain Jacobi integral as

$$\dot{x}^2 + \dot{y}^2 - 2\Omega + c = 0 \qquad (5)$$

where $c$ is integration constant.

## 3 Characteristic equation and its roots

We assume that the infinitesimal body is displaced a little from one of the triangular points $(x_0, y_0)$ to the point $(x_0 + \xi, y_0 + \eta)$ where $\xi$ and $\eta$ are the variation. Hence the equations of motion and the characteristic equation corresponding to Equations (1) will be controlled by

$$\ddot{\xi} - 2n\dot{\eta} = \Omega_{xx}^0 \, \xi + \Omega_{xy}^0 \, \eta \qquad (6.1)$$

$$\ddot{\eta} + 2n\dot{\xi} = \Omega_{xy}^0 \, \xi + \Omega_{yy}^0 \, \eta \qquad (6.2)$$

$$\sigma^4 + \left(4n^2 - \Omega_{xx}^0 - \Omega_{yy}^0\right)\sigma^2 + \Omega_{xx}^0 \Omega_{yy}^0 - (\Omega_{xy}^0)^2 = 0 \qquad (7)$$

where subscripts $x, y$ point to the second partial derivatives of $\Omega$, while superscript 0 indicates that these derivatives are calculated at the triangular points. Therefore, the derivatives are

$$\Omega_{xx}^0 = \frac{3}{4}\left\{\begin{array}{l} 1 + \frac{9}{2}\left(1 - \frac{8}{9}\mu\right)A_1 + \frac{1}{2}\left(1 + 8\mu\right)A_2 - \frac{2}{3}\left(1 - 3\mu\right)p_1 \\[2mm] + \frac{4}{3}\left(1 - \frac{3}{2}\mu\right)p_2 + \frac{5\left(2r_c - 1\right)M_b}{3\left(T^2 + r_c^2\right)^{3/2}} + \frac{4\left(1 - 2\mu\right)^2 M_b}{3\left(T^2 + r_c^2\right)^{5/2}} \end{array}\right\} \qquad (8.1)$$

$$\Omega_{xy}^0 = \pm\frac{3\sqrt{3}}{4}\left(\begin{array}{l} 1 - 2\mu + \frac{19}{6}\left(1 - \frac{26}{19}\mu\right)A_1 + \frac{7}{6}\left(1 - \frac{26}{7}\mu\right)A_2 - \frac{2}{9}\left(1 + \mu\right)p_1 \\[2mm] + \frac{4}{9}\left(1 - \frac{1}{2}\mu\right)p_2 + \frac{11\left(1 - 2\mu\right)\left(2r_c - 1\right)M_b}{9\left(T^2 + r_c^2\right)^{3/2}} + \frac{\left(1 - 2\mu\right)M_b}{\left(T^2 + r_c^2\right)^{5/2}} \end{array}\right)$$

$$(8.2)$$

$$\Omega_{yy}^{0} = \frac{9}{4}\left\{ \begin{array}{l} 1 + \dfrac{11}{6}A_1 + \dfrac{11}{6}A_2 + \dfrac{2}{9}\left(1 - 3\mu\right)p_1 - \dfrac{4}{9}\left(1 - \dfrac{3}{2}\mu\right)p_2 \\[2ex] + \dfrac{7M_b(2r_c - 1)}{9(T^2 + r_c^2)^{3/2}} + \dfrac{M_b}{(T^2 + r_c^2)^{5/2}} \end{array} \right\} \tag{8.3}$$

where $r_c^2 = 1 - \mu + \mu^2$

In the next two subsections, we will present an analytical study about the existence of secular solution and the conditions that enable us to reduce this solution to periodic solution. We establish our main results in theorem 1 and 2.

## 3.1 Existence of secular solution

- **Theorem 1:** The linearized equations of motion of the infinitesimal body in the frame work of the restricted three-body problem has secular solution around the triangular points when $\mu = \mu_c$. Where $\mu_c$ is the critical mass ratio that depends on the parameters of oblateness and radiation pressure for the primaries as well as the effect of gravitational potential from the belt.

where the critical mass given by (Singh and Taura, 2013)

$$\mu_c = \mu_0 + \mu_p + \mu_a + \mu_b$$

$$\mu_0 = \frac{1}{2}\left(1 - \sqrt{\frac{23}{27}}\right)$$

$$\mu_p = -\frac{2\left(p_1 + p_2\right)}{27\sqrt{69}}$$

$$\mu_a = \frac{1}{9}\left(1 - \frac{13}{\sqrt{69}}\right)A_2 - \frac{1}{9}\left(1 + \frac{13}{\sqrt{69}}\right)A_1$$

$$\mu_b = \left( \frac{3}{2} + \frac{\left(76 - 8r_c\right)\left(r_c^2 + T^2\right)}{27\sqrt{69}} - \frac{\left(83 + 12r_c^2\right)}{6\sqrt{69}} \right) \frac{M_b}{\left(r_c^2 + T^2\right)^{5/2}}$$

- **Proof of theorem 1**

If the discriminant of the quadratic in Equation (7) equals zero, the parameter mass $\mu = \mu_c$ (critical mass). Consequently the roots of the characteristic equation can be written as

$$\sigma_{1,2}^2 = -\omega^2 \tag{9.1}$$

and its roots are $\sigma_1 = \sigma_3 = i\,\omega, \quad \sigma_2 = \sigma_4 = -i\,\omega$ where

$$\omega^2 = \frac{1}{2}(4n^2 - \Omega_{xx}^0 - \Omega_{yy}^0) \tag{9.2}$$

Now substituting Equations (4), (8.1) and (8.2) into (9.2), the value of $\omega$ is

$$\omega = \frac{1}{\sqrt{2}} \left( \begin{array}{c} 1 - \dfrac{3}{4}(1-2\mu)A_1 + \dfrac{3}{4}(1-2\mu)A_2 \\[3mm] + \dfrac{M_b(2r_c+3)}{2(T^2+r_c^2)^{3/2}} - \dfrac{13(1 - \dfrac{16}{13}\mu + \dfrac{16}{13}\mu^2)M_b}{8(T^2+r_c^2)^{5/2}} \end{array} \right) \tag{10}$$

The characteristic Equation (7) has four pure imaginary roots and every two of them are equal. Therefore, the solution of Equations (6) contains secular terms. The general solution will be governed by

$$\xi = \left( \alpha_1 + \alpha_2 t \right) cos\omega t + \left( \alpha_3 + \alpha_4 t \right) sin\omega t \tag{11.1}$$

$$\eta = \left( \beta_1 + \beta_2 t \right) cos\omega t + \left( \beta_3 + \beta_4 t \right) sin\omega t \tag{11.2}$$

which ends the proof.

It is clear from Equation (10) that the angular velocity does not affect directly by the radiating forces. Now we show graphically the effect of oblateness and the gravitational potential from a belt on the angular frequency when $T = 0.01$, $M_b = \{0.01, \ 0.02, \ 0.03\}$ and $\mu \in [0, 0.3]$, see Figures (2,3,4 and 5).

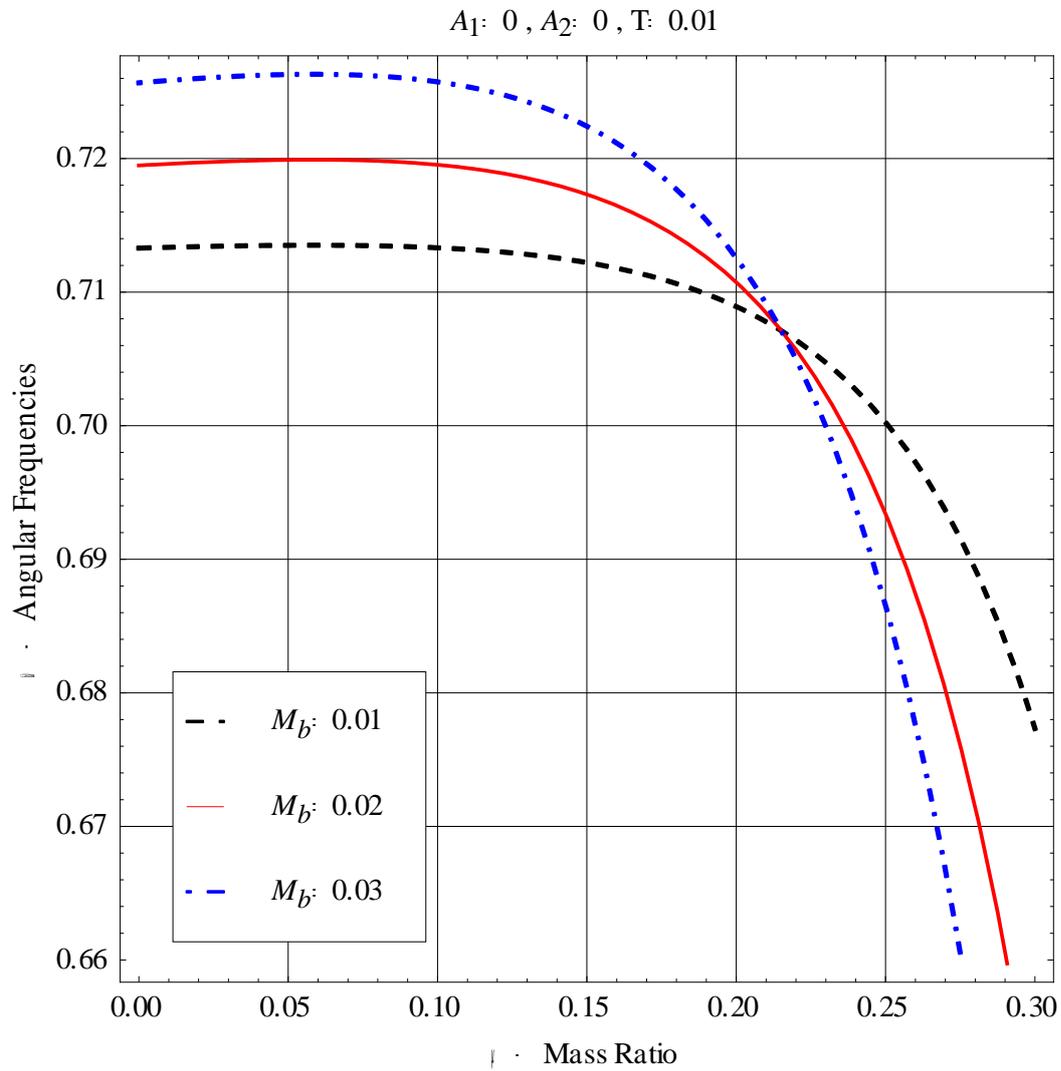

Fig 2: The angular frequency versus the mass ratio $\mu$ when $T = 0.01$, $M_b = \{0.01, \ 0.02, \ 0.03\}$, $A_1 = 0$ and $A_2 = 0$.

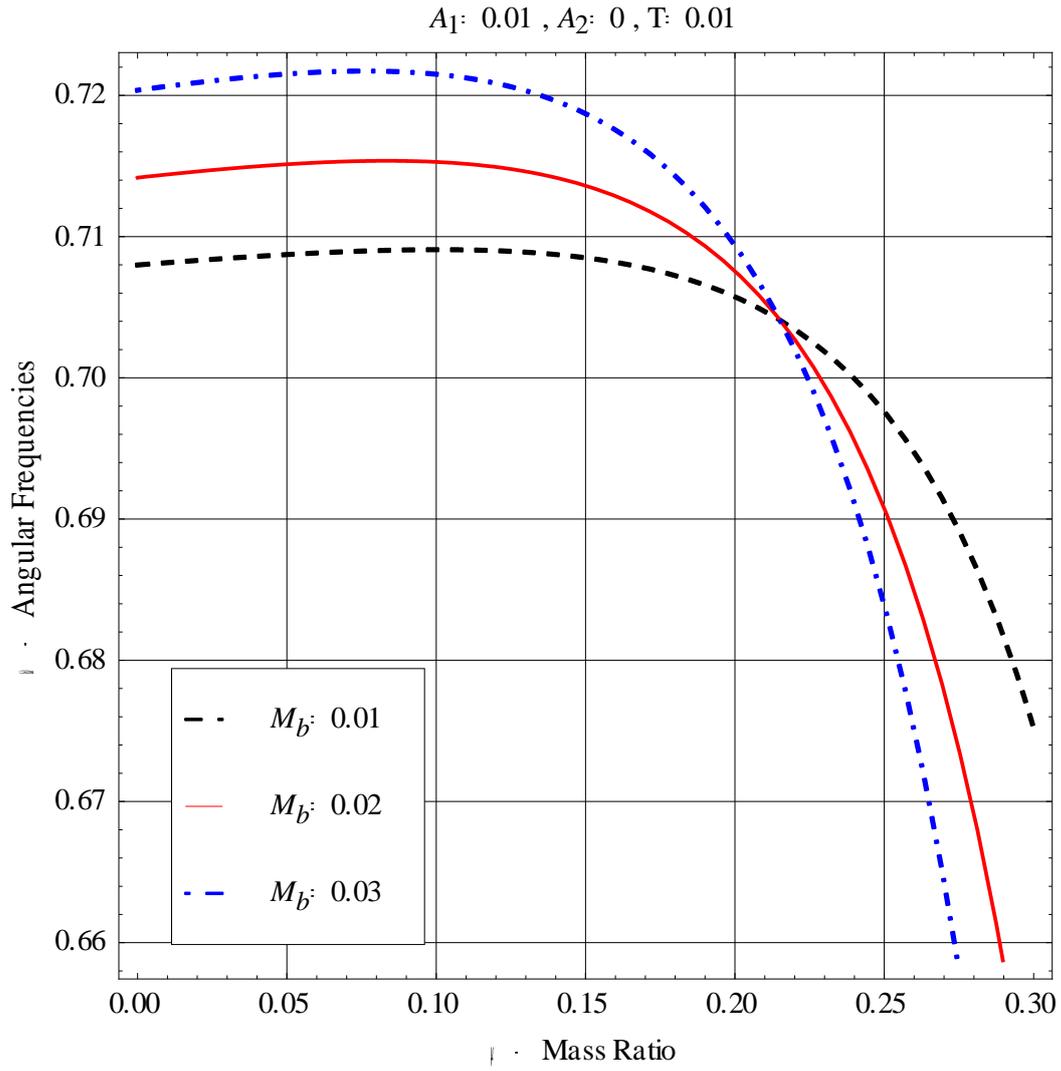

Fig 3: The angular frequency versus the mass ratio $\mu$ when $T = 0.01$, $M_b = \{0.01,\ 0.02,\ 0.03\}$, $A_1 = 0.01$ and $A_2 = 0$

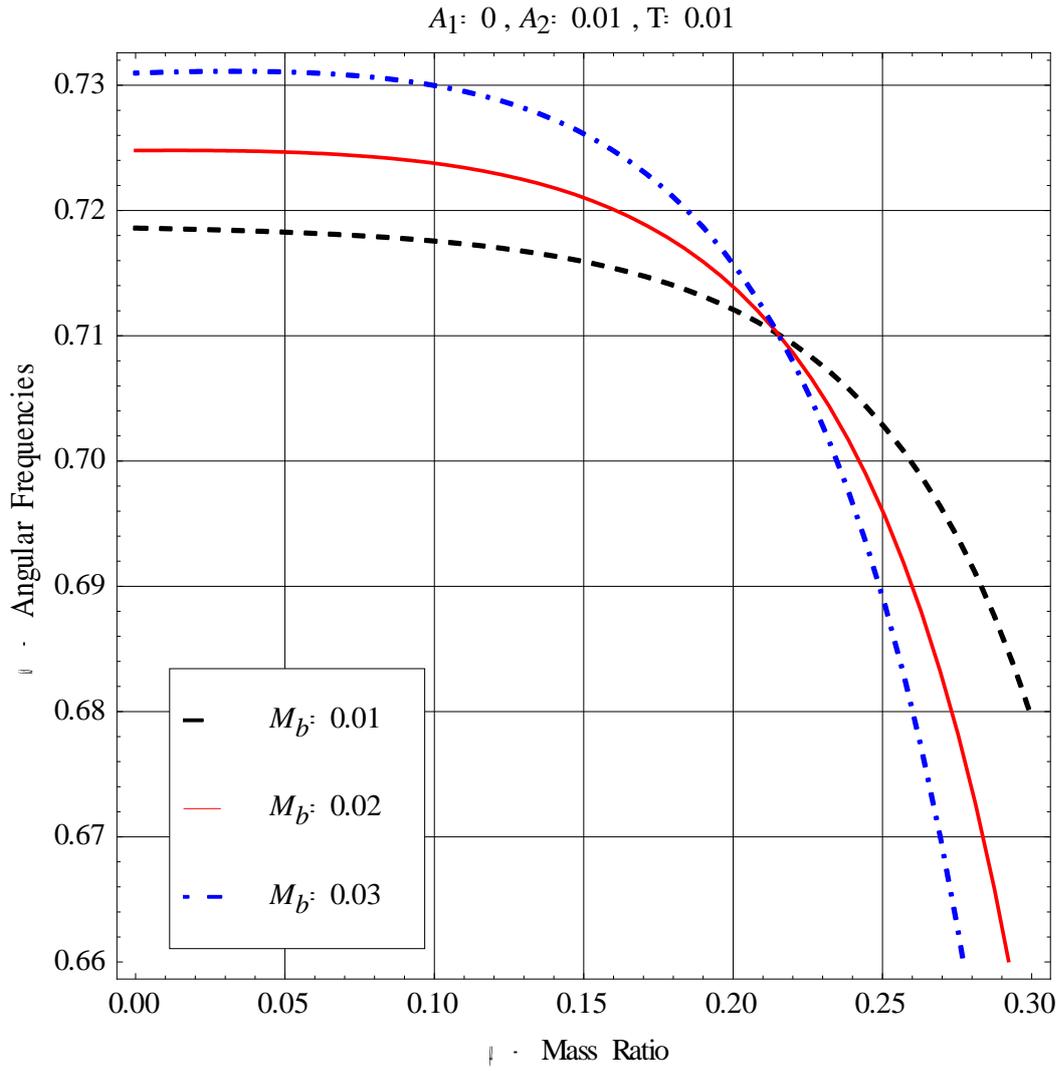

Fig 4: The angular frequency versus the mass ratio $\mu$ when $T = 0.01$, $M_b = \{0.01,\ 0.02,\ 0.03\}$, $A_1 = 0$ and $A_2 = 0.01$.

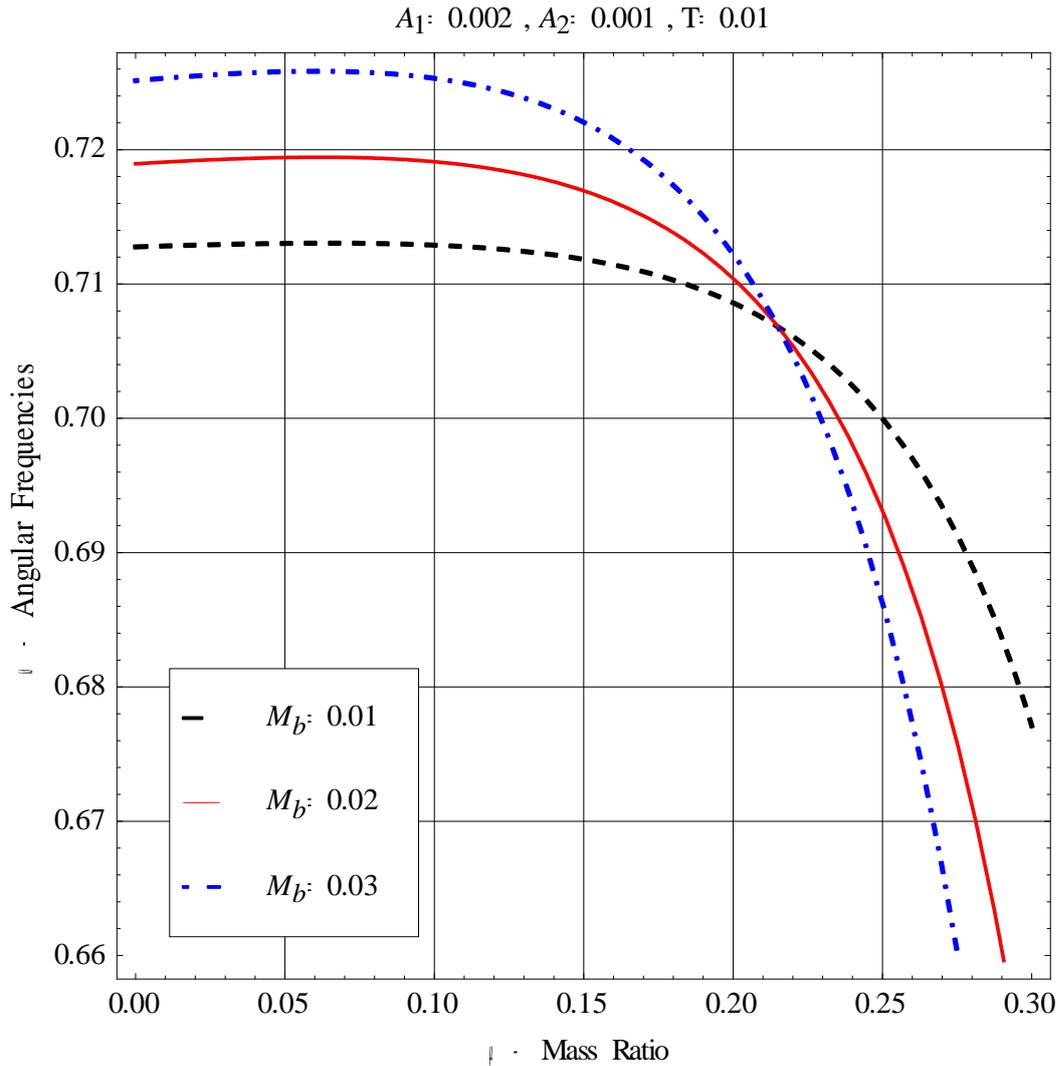

Fig 5: The angular frequency versus the mass ratio $\mu$ when $T = 0.01$, $M_b = \{0.01,\ 0.02,\ 0.03\}$, $A_1 = 0.002$ and $A_2 = 0.001$.

The above figures show that the angular frequency decreases with increment in the mass ratio $\mu$, but the rate of decline is increasing faster whenever the total mass of the belt $M_b$ is increasing.

### 3.2 Existence of Periodic solution

- **Theorem 2:** The secular solution around the triangular equilibrium points can be reduced to periodic solution, when the components of the initial velocities are selected properly.

- **Proof of theorem 2**

Substituting Equations (11.1), (11.2) into (6.1), (6.2) and equating the coefficients of sine and cosine terms respectively, the relations among the coefficients occurring in the solution are

$$\beta_1 = -\gamma \left[ 2\omega\gamma\Omega_{xy}^0\alpha_4 - 2n\omega\alpha_3 - 2n\left(1 - 2\omega^2\gamma\right)\alpha_2 + \Omega_{xy}^0\alpha_1 \right] \qquad (12.1)$$

$$\beta_2 = \gamma \left[ 2n\omega\alpha_4 - \Omega_{xy}^0\alpha_2 \right] \qquad (12.2)$$

$$\beta_3 = \gamma \left[ 2n\left(1 - 2\omega^2\gamma\right)\alpha_4 - \Omega_{xy}^0\alpha_3 - 2\omega\gamma\Omega_{xy}^0\alpha_2 - 2n\omega\alpha_1 \right] \qquad (12.3)$$

$$\beta_4 = -\gamma \left[ \Omega_{xy}^0\alpha_4 + 2n\omega\alpha_2 \right] \qquad (12.4)$$

where

$$\gamma = \frac{\omega^2 + \Omega_{xx}^0}{4n^2\omega^2 + (\Omega_{xy}^0)^2} = \frac{1}{\omega^2 + \Omega_{yy}^0} \qquad (13)$$

We assume that ($\xi_o$, $\eta_0$, $\dot{\xi}_0$ and $\dot{\eta}_0$ ) are the initial conditions at the initial time ($t_0 = 0$). Substituting these values into Equations (11) with using Equations (12), the coefficients of the solution ($\alpha_1$, $\alpha_2$, $\alpha_3$, $\alpha_4$) are related to these condition by the following system

$$A\alpha = C \qquad (14)$$

Where

$$A = \begin{pmatrix} 0 & 0 & 0 & 1 \\ 0 & \omega & 1 & 0 \\ 2\omega\gamma^2\Omega_{xy}^0 & -2n\omega & -2n\left(1 - 2\omega^2\gamma\right) & \gamma\ \Omega_{xy}^0 \\ 4n\omega\gamma\left(1 - \omega^2\gamma\right) & -\omega\gamma\ \Omega_{xy}^0 & -\gamma\ \Omega_{xy}^0\left(1 - 2\omega^2\gamma\right) & -2n\omega^2\gamma \end{pmatrix} \qquad (15.1)$$

$$\alpha = \begin{pmatrix} \alpha_1 \\ \alpha_2 \\ \alpha_3 \\ \alpha_4 \end{pmatrix} \qquad (15.2)$$

$$C = \begin{pmatrix} \xi_0 \\ \dot{\xi}_0 \\ -\eta_0 \\ \dot{\eta}_0 \end{pmatrix} \tag{15.3}$$

By solving the system of Equations (14), the coefficients ($\alpha_1$, $\alpha_2$, $\alpha_3$, $\alpha_4$) will be governed by

$$\alpha_1 = \xi_0 \tag{16.1}$$

$$\alpha_2 = \gamma\dot{\xi}_0 + \frac{\left[\gamma\Omega_{xy}^0\dot{\eta}_0 + 2n\gamma\Omega_{xy}^0\xi_0 + 2n\left(1-\omega^2\gamma\right)\eta_0\right]}{2\omega^2\gamma^2\left[\gamma\left(\Omega_{xy}^0\right)^2 - 4n^2\left(1-\omega^2\gamma\right)\right]} \tag{16.2}$$

$$\alpha_3 = \frac{\begin{bmatrix} \gamma\left(1-\omega^2\gamma\right)\left[4n^2\left(1-\omega^2\gamma\right) - \gamma\left(\Omega_{xy}^0\right)^2\right]\dot{\xi}_0 \\ -\gamma\Omega_{xy}^0\dot{\eta}_0 - 2n\left(1-\omega^2\gamma\right)\eta_0 - 2n\gamma\Omega_{xy}^0\xi_0 \end{bmatrix}}{2\omega^3\gamma^2\left[\gamma\left(\Omega_{xy}^0\right)^2 - 4n^2\left(1-\omega^2\gamma\right)\right]} \tag{16.3}$$

$$\alpha_4 = -\frac{\left[2n\dot{\eta}_0 + \Omega_{xy}^0\eta_0 + \gamma\left(4n^2\omega^2 + \left(\Omega_{xy}^0\right)^2\right)\xi_0\right]}{2\omega\gamma\left[\gamma\left(\Omega_{xy}^0\right)^2 - 4n^2\left(1-\omega^2\gamma\right)\right]} \tag{16.4}$$

With special choice for the constants $\alpha_2$ and $\alpha_4$ ($\alpha_2 = \alpha_4 = 0$), the secular term may be eliminated. Therefore, the components of the initial velocities $\dot{\xi}_0$ and $\dot{\eta}_0$ are controlled by

$$\dot{\xi}_0 = \frac{\left[\eta_0 + \gamma\Omega_{xy}^0\xi_0\right]}{4n\omega^2\gamma^3} \tag{17.1}$$

$$\dot{\eta}_0 = -\frac{1}{2n}\left[\Omega_{xy}^0\eta_0 + \gamma\left(4n^2\omega^2 + \left(\Omega_{xy}^0\right)^2\right)\xi_0\right] \tag{17.2}$$

Consequently Equations (16) can be written in the form

$$\alpha_1 = \xi_0 \tag{18.1}$$

$$\alpha_2 = 0 \tag{18.2}$$

$$\alpha_3 = -\frac{1}{2\omega}\Big[\eta_0 + \gamma\Omega_{xy}^0\xi_0\Big]\Big[1-\omega^2\gamma(1+2\gamma)\Big] \tag{18.3}$$

$$\alpha_4 = 0 \tag{18.4}$$

Substituting Equations (18) into (12) we obtain

$$\beta_1 = \gamma\left(\begin{array}{c} n\Big[1-\omega^2\gamma(1+2\gamma)\Big]\eta_0 \\ -\Omega_{xy}^0\Big[(1-n\gamma)+n\omega^2\gamma^2(1+2\gamma)\Big]\xi_0 \end{array}\right) \tag{19.1}$$

$$\beta_2 = 0 \tag{19.2}$$

$$\beta_3 = -\frac{\gamma}{2\omega}\left(\begin{array}{c} \Omega_{xy}^0\Big[1-\omega^2\gamma(1+2\gamma)\Big]\eta_0 \\ +\Big[\gamma\big(\Omega_{xy}^0\big)^2\Big[1-\omega^2\gamma(1+2\gamma)\Big]+4n\omega^2\Big]\xi_0 \end{array}\right) \tag{19.3}$$

$$\beta_4 = 0 \tag{19.4}$$

Hence the solution can be written as

$$\xi = \alpha_1 cos\omega t + \alpha_3 sin\omega t \tag{20.1}$$

$$\eta = \beta_1 cos\omega t + \beta_3 sin\omega t \tag{20.2}$$

Ending the proof.

- **Remark**

However the triangular points are unstable when the solution contains secular terms. Equations (20) show that the trajectory of the infinitesimal body around this points is elliptical orbit, when both primaries are oblate and radiating, together with the effect of gravitational potential from a belt

Let consider the origin of the coordinates system is one of the triangular points and the infinitesimal body begins its motion from the origin. Therefore the initial conditions are $(\xi_0, \eta_0) = (-x_0, -y_0)$ where $x_0$ and $y_0$ are controlled by (Singh and Taura, 2013)

$$x_0 = \frac{1}{2}\left(1 - 2\mu - \frac{2}{3}(p_1 - p_2) + (A_1 - A_2)\right) \tag{21.1}$$

$$y_0 = \pm\frac{\sqrt{3}}{2}\left(1 - \frac{2}{9}(p_1 + p_2) - \frac{1}{3}(A_1 + A_2) - \frac{4(2r_c - 1)M_b}{9\left(T^2 + r_c^2\right)^{3/2}}\right) \tag{21.2}$$

## 3.3 Ellipse elements

After elimination the time from Equations (20) then they can be written as one equation in the form

$$\left(\beta_1^2 + \beta_3^2\right)\xi^2 - 2\left(\alpha_1\beta_1 + \alpha_3\beta_3\right)\xi\eta + \left(\alpha_1^2 + \alpha_3^2\right)\eta^2 = \left(\alpha_1\beta_3 - \alpha_3\beta_1\right)^2$$

$$\tag{22}$$

Since

$$H = \begin{vmatrix} a_{11} & a_{12} \\ a_{12} & a_{22} \end{vmatrix} > 0 \ ,$$

where

$$a_{11} = \left(\beta_1^2 + \beta_3^2\right), \ a_{12} = -\left(\alpha_1\beta_1 + \alpha_3\beta_3\right) \text{ and } a_{22} = \left(\alpha_1^2 + \alpha_3^2\right)$$

Therefore, Equation (22) represent an ellipse, furthermore this equation does not contain the linear terms of $\xi$ and $\eta$ , then center of ellipse is the origin of the coordinate system $(\xi, \eta)$ which coincides with one of the triangular points $L_4$ and $L_5$. On the other hand, the bilinear term in this equation indicate to the principal axes of the ellipse are rotated around the perpendicular line on the plane ξ-η at origin by an angle $\theta$ , see Figure (6) for details. This observation lead to the idea of introducing the new variables $(u, v)$ such that the bilinear term does not appear again in Equation (22). This variable will be performed by the transformation

$$\xi = ucos\theta - vsin\theta \tag{23.1}$$

$$\eta = u\sin\theta + v\cos\theta \tag{23.2}$$

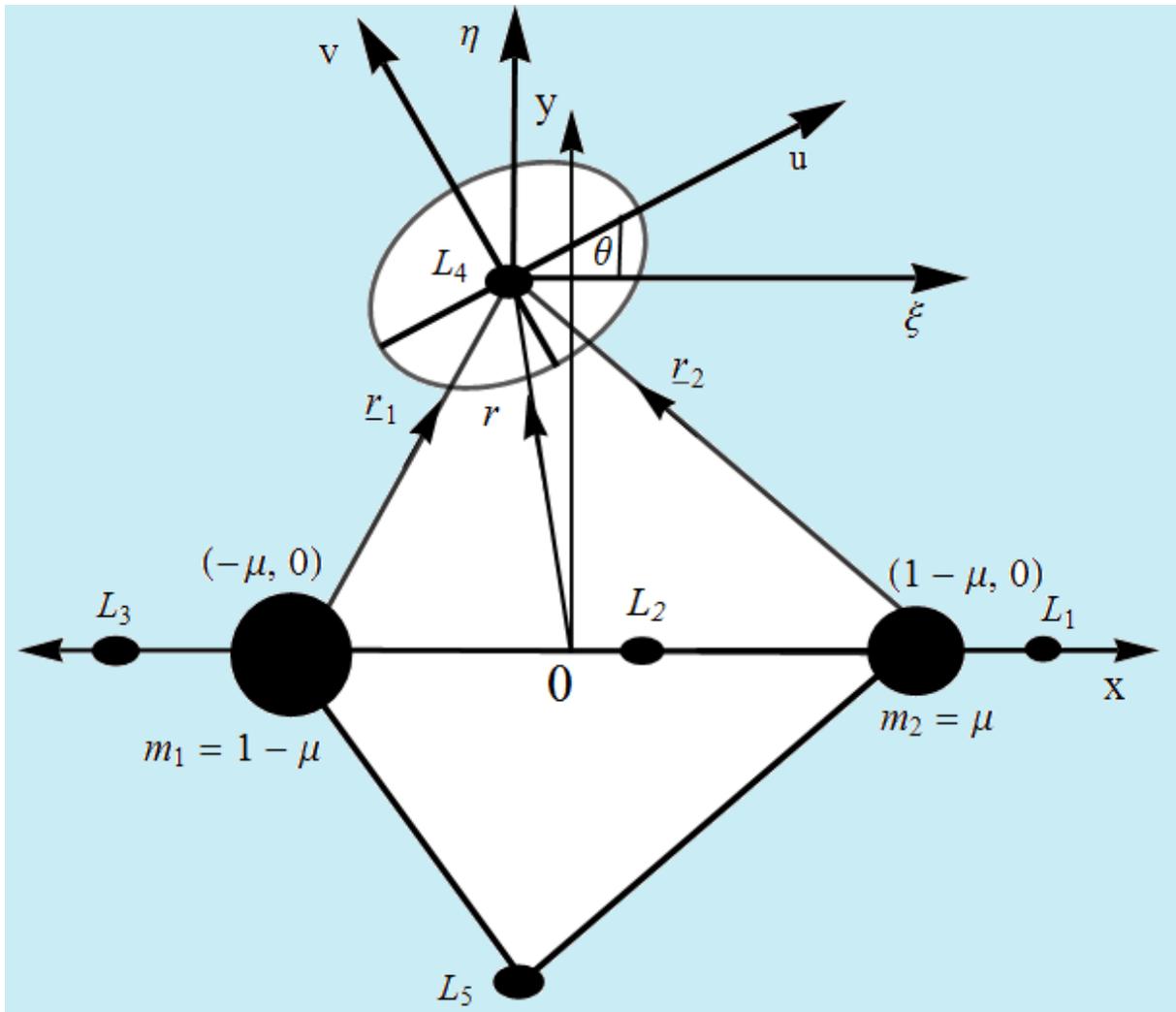

Fig. 6: The configuration of inertial XY and synodic x y coordinate frames of the restricted three-body problem when the primaries are surrounded by the belt of asteroids.

After substituting Equation (23) into (22) with some simple calculations, one obtains

$$u = \alpha \cos \omega t \tag{24.1}$$

$$v = \beta \sin \omega t \tag{24.2}$$

where

$$\alpha^2 = \frac{2(\alpha_1 \beta_3 - \alpha_3 \beta_1)^2}{\left( \begin{pmatrix} \alpha_1^2 + \beta_1^2 \\ + \alpha_3^2 + \beta_3^2 \end{pmatrix} - \begin{pmatrix} \alpha_1^2 - \beta_1^2 \\ + \alpha_3^2 - \beta_3^2 \end{pmatrix} \cos 2\theta - 2(\alpha_1 \beta_1 + \alpha_3 \beta_3) \sin 2\theta \right)}$$

$$\tag{25.1}$$

$$\beta^2 = \frac{2(\alpha_1 \beta_3 - \alpha_3 \beta_1)^2}{\left( \begin{pmatrix} \alpha_1^2 + \beta_1^2 \\ + \alpha_3^2 + \beta_3^2 \end{pmatrix} + \begin{pmatrix} \alpha_1^2 - \beta_1^2 \\ + \alpha_3^2 - \beta_3^2 \end{pmatrix} \cos 2\theta + 2(\alpha_1 \beta_1 + \alpha_3 \beta_3) \sin 2\theta \right)}$$

$$\tag{25.2}$$

$$\tan 2\theta = \frac{2(\alpha_1 \beta_1 + \alpha_3 \beta_3)}{\left( \alpha_1^2 - \beta_1^2 + \alpha_3^2 - \beta_3^2 \right)} \tag{26}$$

Here $\alpha$ and $\beta$ denote the lengths of semi-major and semi-minor axes of the ellipse respectively, while $\theta$ gives the direction of the principals axes. In addition, the eccentricity $e$ and the period of motion $\tau$ are

$$e^2 = \frac{2\left[ \left( \alpha_1^2 - \beta_1^2 + \alpha_3^2 - \beta_3^2 \right) \cos \theta + 2(\alpha_1 \beta_1 + \alpha_3 \beta_3) \sin 2\theta \right]}{\left[ \begin{pmatrix} \alpha_1^2 + \beta_1^2 \\ + \alpha_3^2 + \beta_3^2 \end{pmatrix} + \begin{pmatrix} \alpha_1^2 - \beta_1^2 \\ + \alpha_3^2 - \beta_3^2 \end{pmatrix} \cos 2\theta + 2(\alpha_1 \beta_1 + \alpha_3 \beta_3) \sin 2\theta \right]} \tag{27}$$

$$\tau = \frac{2\pi}{\omega} \qquad\qquad (28)$$

## 4 Numerical Applications

In this section we show numerically and graphically the effects of the perturbed forces on the ellipse elements that describe the pass of the infinitesimal body around the triangular equilibrium point $L_4$. To facilitate our calculus, firstly we will introduce an algorithm to determine the position of the infinitesimal body at any time. This algorithm will be performed in the following steps:

- **Algorithm**

  Start

1) Determine the parameters $\mu$ , $A_1$ , $A_2$ , $p_1$ , $p_2$ , $M_b$ and $T$ for any given system.
2) Evaluate $n$ , $r_c$ , $\Omega_{xx}^0$, $\Omega_{xy}^0$ and $\Omega_{yy}^0$ .
3) Evaluate $\omega$
4) Evaluate $\gamma$
5) Evaluate $\xi_0$ and $\eta_0$ .
6) Evaluate $\alpha_1, \alpha_3$ , $\beta_1$ and $\beta_3$ .
7) Evaluate $\cos 2\theta$ and $\sin 2\theta$ .
8) Evaluate $\alpha$ and $\beta$ .
9) Use steps (3) and (8) to find the coordinates of the infinitesimal body at any time.

   End.

Now, the values of the lengths of the semi-major and the semi-minor axes, the eccentricity, the angular velocity and the period of motion ($\alpha$ , $\beta$ , $e$ , $\omega$ and $\tau$ ) under the perturbations effect of the oblateness, the radiating and the gravitational potential from the belt are calculated and tabulated in Tables (1,2,3 and4).

**Table 1:** The effect of the density profile and the total mass of the belt on the parameters that describe the trajectory of the infinitesimal body when $\mu = 0.03$ in the absence of the oblateness and radiation influences.

**Here Table 1**

**Table 2:** The effect of the oblateness on the parameters that describe the trajectory of the infinitesimal body when $\mu = 0.03$, $M_b = 0.04$ and $T = 0.01$ in the absence of the radiation influences.

**Here Table 2**

**Table 3:** The effect of the radiation on the parameters that describe the trajectory of the infinitesimal body when $\mu = 0.03$, $M_b = 0.04$ and $T = 0.01$ in the absence of the oblateness influences.

**Here Table 3**

**Table 4:** The effect of the oblateness and radiation on the parameters that describe the trajectory of the infinitesimal body when $\mu = 0.03$, $M_b = 0.04$ and $T = 0.01$.

**Here Table 4**

After examining the values of the elements of the ellipse, the angular frequency and the period of motion in Table (1, 2, 3 and 4), we find that these values are affected by the changes of perturbations. As a result, the trajectory of the infinitesimal body will be also affected by the perturbed forces. Moreover, we investigate the changes of the trajectory of the infinitesimal body graphically for the first case in every tables, see Figurers (7, 8, 9 and 10)

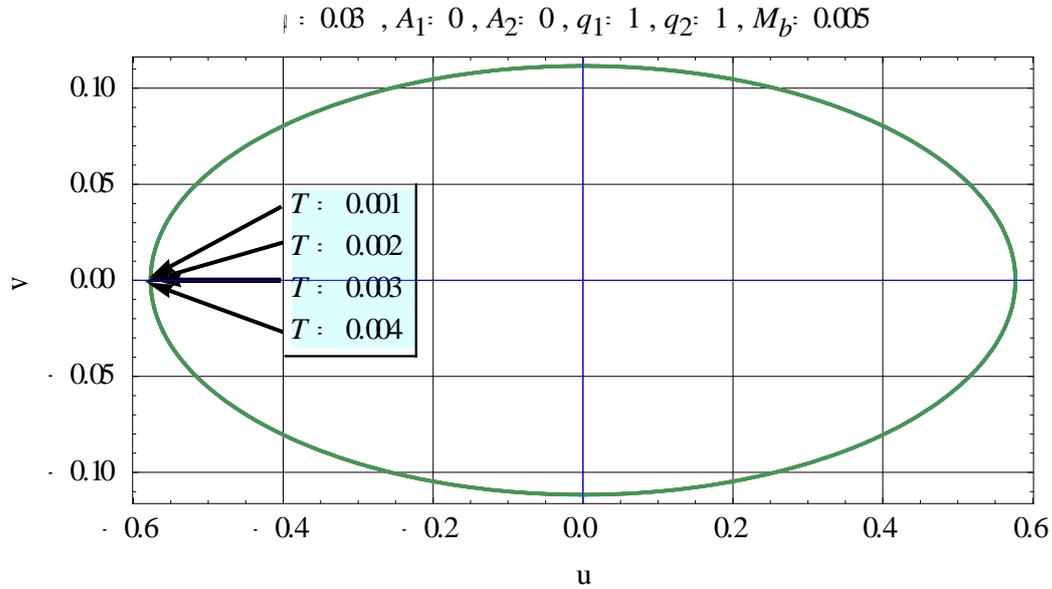

**Fig. 7:** The effect of the density profile of the belt on the trajectory of the infinitesimal body when $\mu = 0.03$ and $M_b = 0.05$ in the absence of the oblateness and radiation influences (Table 1 – Case 1).

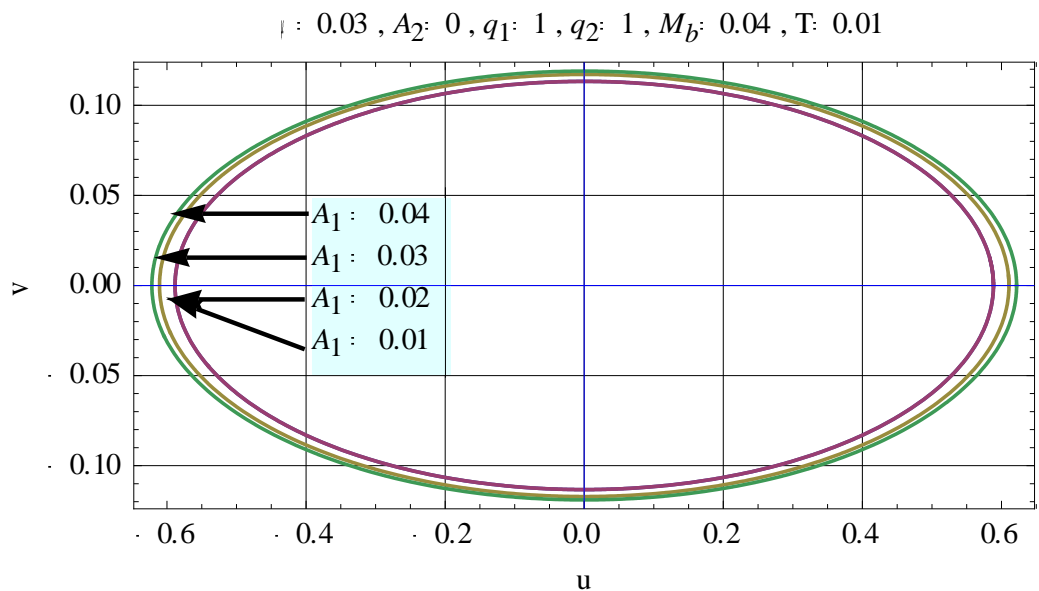

**Fig. 8:** The oblateness effect of the bigger primary on the trajectory of the infinitesimal body when $\mu = 0.03$, $A_2 = 0$, $M_b = 0.04$ and $T = 0.01$ as well as the primaries are no radiating (Table 2 – Case 1).

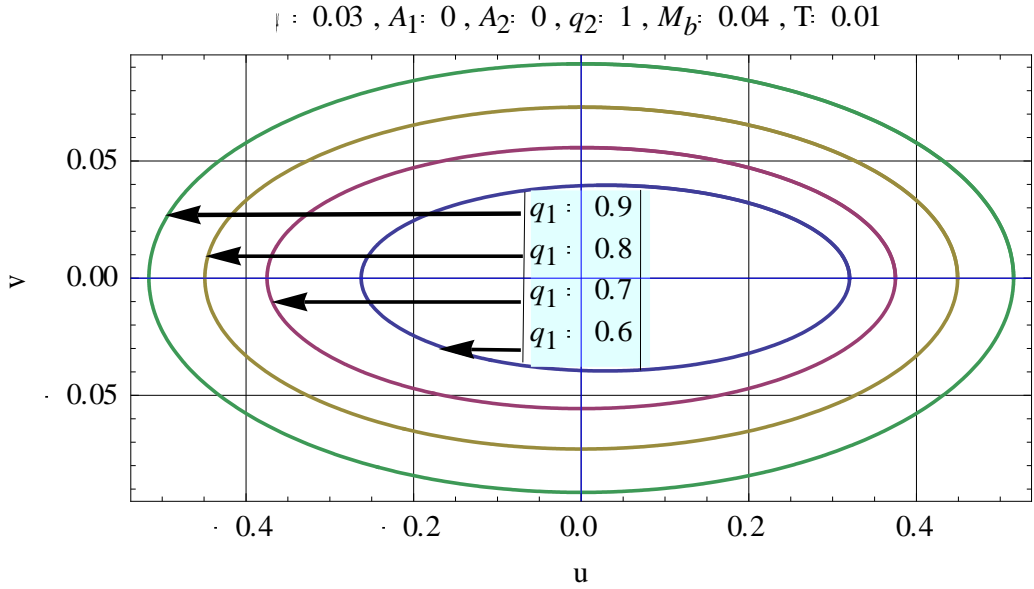

**Fig. 9:** The radiation effect of the bigger primary on the trajectory of the infinitesimal body when $\mu = 0.03$, $q_2 = 0.04$, $M_b = 0.04$ and $T = 0.01$ in the absence the oblateness influences of the primaries (Table 3 – Case 1).

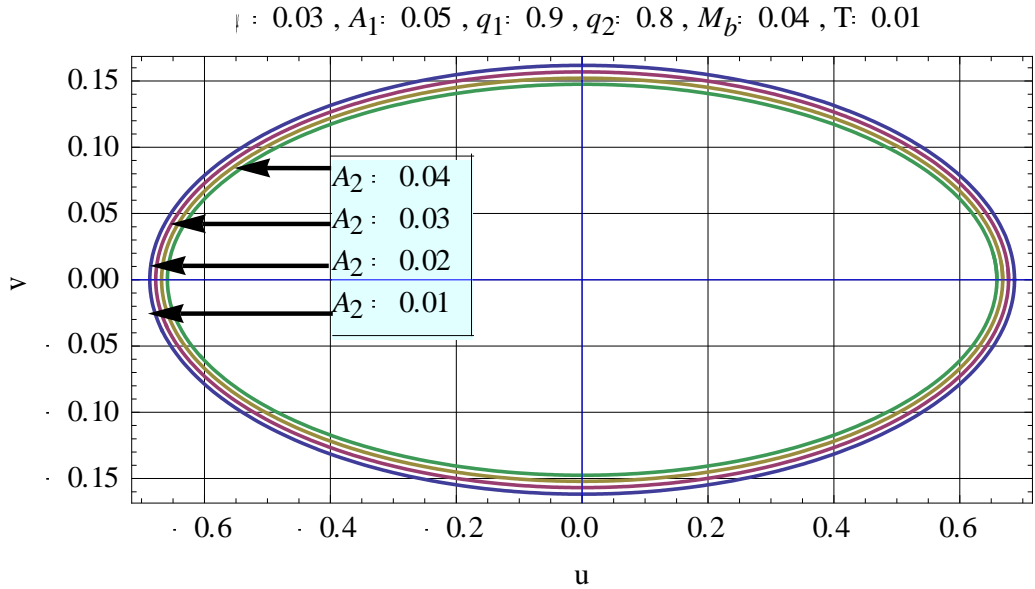

**Fig. 10:** The oblateness effect of the smaller primary on the trajectory of the infinitesimal body when $\mu = 0.03$, $A_1 = 0.05$ $M_b = 0.04$ and $T = 0.01$ as well as both primaries are radiating (Table 4 – Case 1).

## 5  Conclusion

The model of the photogravitational restricted three-body problem is generalized in sense that both the primaries are oblate spheroid as well as the gravitational potential from a belt. We show that the linearized equation of motion of the infinitesimal body in the proximity of the triangular equilibrium points has a secular solution when the mass ratio equals the critical mass value. We also show that this solution can be reduced to periodic solution. The main result of this work are demonstrated in theorems (1 and 2). Furthermore, some numerical and graphical investigations are introduced.

We show graphically the influence of oblateness and the gravitational potential from a belt on the behavior of angular frequency when $T = 0.01$, $M_b = \{0.01, \ 0.02, \ 0.03\}$ and $\mu \in [0, 0.3]$.

Table 1 is devoted to show the parameters effect that characterize the total mass and density profile of the belt when the influence of the oblateness and radiating parameters are switched off as well as the mass ratio is constants ($\mu = 0.03$). We found that the length of semi-major axis is directly proportional to the total mass of the belt while the opposite happens due to the semi-minor axis. The eccentricity, the angular frequency and the period of motion may be affected also according to the changes of total mass of the belt but this changes is very week.

In Table 2, the changes of the ellipse elements are showed due to the oblateness parameters when ($\mu = 0.03$, $M_b = 0.04$ and $T = 0.01$). We found that the semi-axes increase with increments in the parameter that characterize the oblateness of the bigger primary while they decrease due to the increasing in the oblateness parameter of the smaller body.

The semi-major and semi-minor axes are increasing with increment in radiating factor of the bigger primary, while they decrease due to the increasing in radiating factor of the smaller primary. These data is tabulated in Table 3 when ($\mu = 0.03$, $M_b = 0.04$ and $T = 0.01$).

In addition, the effect of the oblateness, radiation forces, the gravitational potential from a belt together on the ellipse elements is demonstrated in Table 4. Moreover, the changes of the trajectory of the infinitesimal body are investigated graphically for the first case in every tables. In general, the parameters ($\alpha$, $\beta$, $e$, $\omega$, $\tau$) that describe the motion of the infinitesimal body may be affected by the perturbed forces that are aforementioned.

## Acknowledgment

The authors (especially the first author) wish to express their gratitude to referees for their useful suggestions and criticism which improved the presentation of the paper.

<div align="center">

**References**

</div>

| Case | $\mu$ | $A_1$ | $A_2$ | $q_1$ | $q_2$ | $M_b$ | $T$ | $\alpha$ | $\beta$ | $e$ | $\omega$ | $\tau$ |
|------|-------|-------|-------|-------|-------|-------|-----|----------|---------|-----|----------|--------|
| 1 | 0.03 | 0 | 0 | 1 | 1 | 0.005 | 0.001 | 0.576883 | 0.111611 | 0.981106 | 0.710287 | 8.84598 |
| | 0.03 | 0 | 0 | 1 | 1 | 0.005 | 0.002 | 0.576883 | 0.111611 | 0.981106 | 0.710287 | 8.84598 |
| | 0.03 | 0 | 0 | 1 | 1 | 0.005 | 0.003 | 0.576883 | 0.111611 | 0.981106 | 0.710287 | 8.84598 |
| | 0.03 | 0 | 0 | 1 | 1 | 0.005 | 0.004 | 0.576883 | 0.111611 | 0.981106 | 0.710287 | 8.84598 |
| 2 | 0.03 | 0 | 0 | 1 | 1 | 0.002 | 0.001 | 0.576856 | 0.111644 | 0.981092 | 0.708379 | 8.86981 |
| | 0.03 | 0 | 0 | 1 | 1 | 0.003 | 0.001 | 0.576865 | 0.111633 | 0.981097 | 0.709015 | 8.86185 |
| | 0.03 | 0 | 0 | 1 | 1 | 0.004 | 0.001 | 0.576874 | 0.111622 | 0.981101 | 0.709651 | 8.85391 |
| | 0.03 | 0 | 0 | 1 | 1 | 0.005 | 0.001 | 0.576883 | 0.111611 | 0.981106 | 0.710287 | 8.84598 |
| 3 | 0.03 | 0 | 0 | 1 | 1 | 0.01 | 0.01 | 0.576925 | 0.111559 | 0.981126 | 0.713467 | 8.80655 |
| | 0.03 | 0 | 0 | 1 | 1 | 0.02 | 0.01 | 0.576993 | 0.111467 | 0.981162 | 0.719828 | 8.72873 |
| | 0.03 | 0 | 0 | 1 | 1 | 0.03 | 0.01 | 0.577035 | 0.111388 | 0.981192 | 0.726189 | 8.65228 |
| | 0.03 | 0 | 0 | 1 | 1 | 0.04 | 0.01 | 0.577045 | 0.111320 | 0.981216 | 0.732549 | 8.57715 |

Table 1

| Case | $\mu$ | $A_1$ | $A_2$ | $q_1$ | $q_2$ | $M_b$ | $T$ | $\alpha$ | $\beta$ | $e$ | $\omega$ | $\tau$ |
|------|-------|-------|-------|-------|-------|-------|-----|----------|---------|-----|----------|--------|
| 1 | 0.03 | 0.01 | 0 | 1 | 1 | 0.04 | 0.01 | 0.588395 | 0.113308 | 0.981283 | 0.727564 | 8.63592 |
|  | 0.03 | 0.02 | 0 | 1 | 1 | 0.04 | 0.01 | 0.599620 | 0.115255 | 0.981353 | 0.722579 | 8.69550 |
|  | 0.03 | 0.03 | 0 | 1 | 1 | 0.04 | 0.01 | 0.610716 | 0.117159 | 0.981427 | 0.717594 | 8.75591 |
|  | 0.03 | 0.04 | 0 | 1 | 1 | 0.04 | 0.01 | 0.621681 | 0.119018 | 0.981503 | 0.712609 | 8.81716 |
| 2 | 0.03 | 0 | 0.01 | 1 | 1 | 0.04 | 0.01 | 0.566818 | 0.107374 | 0.981894 | 0.737534 | 8.51918 |
|  | 0.03 | 0 | 0.02 | 1 | 1 | 0.04 | 0.01 | 0.556687 | 0.103576 | 0.982539 | 0.742519 | 8.46198 |
|  | 0.03 | 0 | 0.03 | 1 | 1 | 0.04 | 0.01 | 0.546642 | 0.099919 | 0.983153 | 0.747504 | 8.40555 |
|  | 0.03 | 0 | 0.04 | 1 | 1 | 0.04 | 0.01 | 0.536676 | 0.096393 | 0.983737 | 0.752490 | 8.34986 |
| 3 | 0.03 | 0.04 | 0.01 | 1 | 1 | 0.04 | 0.01 | 0.611767 | 0.115141 | 0.982129 | 0.717594 | 8.75591 |
|  | 0.03 | 0.04 | 0.02 | 1 | 1 | 0.04 | 0.01 | 0.601953 | 0.111401 | 0.982726 | 0.722579 | 8.69550 |
|  | 0.03 | 0.05 | 0.03 | 1 | 1 | 0.04 | 0.01 | 0.603212 | 0.109640 | 0.983343 | 0.722579 | 8.69550 |
|  | 0.03 | 0.05 | 0.04 | 1 | 1 | 0.04 | 0.01 | 0.593630 | 0.106163 | 0.983879 | 0.727564 | 8.63592 |

Table 2

| Case | $\mu$ | $A_1$ | $A_2$ | $q_1$ | $q_2$ | $M_b$ | $T$ | $\alpha$ | $\beta$ | $e$ | $\omega$ | $\tau$ |
|---|---|---|---|---|---|---|---|---|---|---|---|---|
| 1 | 0.03 | 0 | 0 | 0.6 | 1 | 0.04 | 0.01 | 0.291319 | 0.0396298 | 0.990704 | 0.732549 | 8.57715 |
|  | 0.03 | 0 | 0 | 0.7 | 1 | 0.04 | 0.01 | 0.374869 | 0.0557037 | 0.988898 | 0.732549 | 8.57715 |
|  | 0.03 | 0 | 0 | 0.8 | 1 | 0.04 | 0.01 | 0.449072 | 0.0729587 | 0.986714 | 0.732549 | 8.57715 |
|  | 0.03 | 0 | 0 | 0.9 | 1 | 0.04 | 0.01 | 0.515870 | 0.0914688 | 0.984155 | 0.732549 | 8.57715 |
| 2 | 0.03 | 0 | 0 | 1 | 0.6 | 0.04 | 0.01 | 0.794318 | 0.281839 | 0.934935 | 0.732549 | 8.57715 |
|  | 0.03 | 0 | 0 | 1 | 0.7 | 0.04 | 0.01 | 0.750166 | 0.230892 | 0.951455 | 0.732549 | 8.57715 |
|  | 0.03 | 0 | 0 | 1 | 0.8 | 0.04 | 0.01 | 0.701456 | 0.186015 | 0.964198 | 0.732549 | 8.57715 |
|  | 0.03 | 0 | 0 | 1 | 0.9 | 0.04 | 0.01 | 0.645198 | 0.146382 | 0.973923 | 0.732549 | 8.57715 |
| 3 | 0.03 | 0 | 0 | 0.8 | 0.6 | 0.04 | 0.01 | 0.704350 | 0.219462 | 0.950220 | 0.732549 | 8.57715 |
|  | 0.03 | 0 | 0 | 0.8 | 0.7 | 0.04 | 0.01 | 0.657241 | 0.175835 | 0.963548 | 0.732549 | 8.57715 |
|  | 0.03 | 0 | 0 | 0.9 | 0.7 | 0.04 | 0.01 | 0.704045 | 0.202364 | 0.957801 | 0.732549 | 8.57715 |
|  | 0.03 | 0 | 0 | 0.9 | 0.8 | 0.04 | 0.01 | 0.652781 | 0.160784 | 0.969192 | 0.732549 | 8.57715 |

Table 3

| Case | $\mu$ | $A_1$ | $A_2$ | $q_1$ | $q_2$ | $M_b$ | $T$ | $\alpha$ | $\beta$ | $e$ | $\omega$ | $\tau$ |
|---|---|---|---|---|---|---|---|---|---|---|---|---|
| 1 | 0.03 | 0.05 | 0.01 | 0.9 | 0.8 | 0.04 | 0.01 | 0.686914 | 0.161856 | 0.971844 | 0.712609 | 8.81716 |
|  | 0.03 | 0.05 | 0.02 | 0.9 | 0.8 | 0.04 | 0.01 | 0.677462 | 0.156921 | 0.972804 | 0.717594 | 8.75591 |
|  | 0.03 | 0.05 | 0.03 | 0.9 | 0.8 | 0.04 | 0.01 | 0.668141 | 0.152171 | 0.973719 | 0.722579 | 8.69550 |
|  | 0.03 | 0.05 | 0.04 | 0.9 | 0.8 | 0.04 | 0.01 | 0.658941 | 0.147593 | 0.974592 | 0.727564 | 8.63592 |
| 2 | 0.03 | 0.02 | 0.01 | 0.9 | 0.8 | 0.04 | 0.01 | 0.660772 | 0.158075 | 0.970963 | 0.727564 | 8.63592 |
|  | 0.03 | 0.03 | 0.01 | 0.9 | 0.8 | 0.04 | 0.01 | 0.669534 | 0.159368 | 0.971258 | 0.722579 | 8.69550 |
|  | 0.03 | 0.04 | 0.01 | 0.9 | 0.8 | 0.04 | 0.01 | 0.678249 | 0.160629 | 0.971551 | 0.717594 | 8.75591 |
|  | 0.03 | 0.05 | 0.01 | 0.9 | 0.8 | 0.04 | 0.01 | 0.686914 | 0.161856 | 0.971844 | 0.712609 | 8.81716 |
| 3 | 0.03 | 0.04 | 0.02 | 0.8 | 0.6 | 0.04 | 0.01 | 0.715450 | 0.207362 | 0.957077 | 0.722579 | 8.69550 |
|  | 0.03 | 0.04 | 0.02 | 0.8 | 0.7 | 0.04 | 0.01 | 0.671153 | 0.168863 | 0.967831 | 0.722579 | 8.69550 |
|  | 0.03 | 0.05 | 0.03 | 0.9 | 0.6 | 0.04 | 0.01 | 0.757238 | 0.227030 | 0.953998 | 0.722579 | 8.69550 |
|  | 0.03 | 0.05 | 0.03 | 0.9 | 0.7 | 0.04 | 0.01 | 0.715063 | 0.187505 | 0.965008 | 0.722579 | 8.69550 |

Table 4